\documentclass[preprint2]{aastex}
\usepackage{graphicx}
\usepackage{pictex}
\usepackage{rotating}

\shorttitle{spectroscopy of Cepheus\,E}
\shortauthors{Smith et al.}

\begin{document}
 
\title{Multi-wavelength spectroscopy of the bipolar outflow from
Cepheus\,E
\thanks{Based on observations with ISO, an ESA project with instruments funded
by ESA Member States (especially the PI countries: France, Germany, the
Netherlands and the United Kingdom) and with the participation of ISAS and
NASA.}}

\author{Michael D. Smith}
\affil{Armagh Observatory, College Hill, Armagh BT61 9DG,
       Northern Ireland}

\author{Dirk Froebrich, Jochen Eisl\"offel}
\affil{Th\"uringer Landessternwarte Tautenburg, Sternwarte 5, 
       D-07778 Tautenburg, Germany}

\begin{abstract} 
Cepheus~E is the site of an exceptional example of a protostellar outflow with
a very young dynamical age and extremely high near infrared luminosity. We
combine molecular spectroscopic data from the submillimeter to the near
infrared in order to interpret the rotational excitation of CO and the
ro-vibrational excitation of H$_2$. We conclude that C-type shocks with a
paraboloidal bow shock geometry can simultaneously explain all the molecular
excitations. Extinction accounts for the deviation of the column densities from
local thermodynamic equilibrium.  A difference in the extinction between the
red and blue-shifted outflow lobes may account for the  measured flux
difference. The outflow is deeply embedded in a clump of density
10$^5$\,cm$^{-3}$,  yet a good fraction of atomic hydrogen, about 40\%, is
required to explain the excitation and statistical equilibrium. We propose that
this atomic component arises, self-consistently, from the dissociated gas at
the apex of the leading bow shocks and the relatively long molecule reformation
time. At least 20 bow shocks are required in each lobe, although these may be
sub-divided into smaller bows and turbulent shocked regions.  The total outflow
mechanical power and cooling amounts to over 30\,L$_{\odot}$, almost half the
source's bolometric luminosity. Nevertheless, only about 6\% of the clump  mass
has been set in outward motion by the outflow, allowing a collapse to
continue.  
\end{abstract}



\section{Introduction}

Twin collimated lobes  of molecular gas stream away from newly forming  stars
\citep{1996ARA&A..34..111B}. These bipolar outflows possess particularly  high
power and thrust during the main phase of inflow onto the protostar.   The
driving agents are often recognised as pulsating supersonic jets, originating
from  near the protostellar surface. The extended environment is pushed and 
shocked, producing bow-shaped structures called Herbig-Haro (HH) objects
\citep{2001ARA&A..39..403R}. The HH objects are often optically invisible
because  the protostar is deeply enshrouded in a dusty cloud. We analyse here
one such outflow, originating from the Class\,0 source Cepheus\,E\,--\,MM
\citep{lel96},  through combined near-infrared data, mid- and far-infrared
(FIR) ISO spectra and  sub-millimeter maps. The numerous emission line
strengths at these long wavelengths  constrain and relate the gas components. 

Cepheus\,E contains a powerful outflow from a luminous source.  A CO-derived
kinetic luminosity of 0.2\,L$_\odot$ \citep{mnmtcs01}, a far infrared line
luminosity of 2.8\,L$_\odot$ \citep{gnl01} and an H$_2$  ro-vibrational line
luminosity of 0.7\,L$_\odot$ \citep{fshe03} have been  derived, based on simple
assumptions and an estimated distance of  $\sim$\,730\,pc. The driving
protostar has a bolometric luminosity of  $\sim$ 80\,L$_\odot$
\citep{lel96,fshe03} and is surrounded by a protostellar  envelope of
25\,M$_\odot$ \citep{lel96,cwknrs01}. This suggests that the  driving young
stellar object will develop into an intermediate mass object,  consistent with
a straightforward evolutionary model \citep{fshe03}. This  contrasts with other
well-known text-book examples of jet-driven Class\,0  outflows, such as HH\,211
and HH\,212, which may be powered by low-mass or  solar-mass stars. These
statements are, however, based on evolutionary  assumptions which need to be
tested.

The outflow is apparently driven by jets or bullets with radial velocities of
-120 and +80\,km\,s$^{-1}$ \citep{1997A&A...323..223S}, as derived from CO
spectra \citep{lel96,hfl99}. The proper motion of an optical knot at the edge
of the southern blue-shifted lobe, HH\,377, is 107$\pm$14\,km\,s$^{-1}$
\citep{ng01} with a radial velocity of -70$\pm$10\,km\,s$^{-1}$
\citep{angcrbr00}. This implies that the jet is more dense than the environment
through which it propagates. Reversing the well-known  formula for thrust
balance yields a jet-ambient density ratio of $\eta = 1/({\rm v}_{jet}/{\rm
v}_{bow} - 1)^2$ = 2.0 (taking the radial velocity components and assuming the
gas ahead of the bow shock is stationary). The inclination angle to the
line-of-sight is  $\sim$\,tan$^{-1}\,(107/70)$ = 57$^\circ\pm$7$^\circ$ and the
bow speed is  128$\pm$12\,km\,s$^{-1}$.

We feature Cep\,E in this study because of its extreme youth and high
luminosity. Several other properties have attracted attention to this outflow.

\begin{itemize} \item The low excitation. Both near-infrared H$_2$ and  optical
H lines yield remarkably low excitation. Despite the strong emission fluxes
stemming from vibrationally excited H$_2$ states, the excitation temperature is
low, with its precise value sensitive to the excitation levels between which it
is measured \citep{esdr96}. The shock speed of HH~377 derived from H$\beta$ and
H$\alpha$ fluxes are under  20~km~s$^{-1}$ \citep{angcrbr00}. \item The
constant molecular excitation.  The H$_2$ line ratios vary remarkably little
over the entire outflow on small and large scales \citep{esdr96}.\\ \item  The
evidence for a precessing underlying flow \citep{esdr96}\\ \end{itemize}

We here analyse the radiative shock waves, {\em simultaneously} modelling the
many strong emission lines from carbon monoxide and hydrogen molecules  through
both rotational and ro-vibrational  transitions. To accomplish this we combine
ISO and ground-based data. We thus hope to go a stage further than previous
studies of  K-band spectra and ISO data, analysed in isolation. Secondly, our
model includes bow shock dynamics and ambipolar diffusion (C-shock) physics, both
of which have not yet been applied to recent Cepheus E spectroscopic data
(although \citet{1997ApJ...474..749L} compared Smith's (1995) tabulated C and
J-type  planar models to K-band spectra). 

Using refurbished shock codes, tested by  \citet{fse02a} for the outflows from
Cep\,A and L\,1448, we model the outflow as planar and bow shocks with a
symmetric shape Z\,$\propto$\,R$^s$ (in cylindrical coordinates). The main aim
of this process is to draw conclusions about the physics, chemistry and
dynamics of the shocks and the properties of the surrounding gas. This analysis
will also provide hints concerning the  early evolution of an outflow and
possibly of the source itself.

In Sect.\,\ref{data} we describe the observations and data reduction. The main
results are presented in Sect.\,\ref{results}. The results of the extinction
modelling (Sect.\,\ref{extinct}), and modelling of the H$_2$ and CO line fluxes
(Sect.\,\ref{models}) are then given.

\section{Observations and Data Reduction}     
\label{data}

\subsection{Near infrared data}

\begin{figure}[t]
\beginpicture
\setcoordinatesystem units <8.5mm,8.5mm> point at 0 0
\setplotarea x from 1 to 9 , y from 1 to 10.9
\put {\resizebox{63.75mm}{!}{
\includegraphics[angle=-90,bb=62 127 517 489]{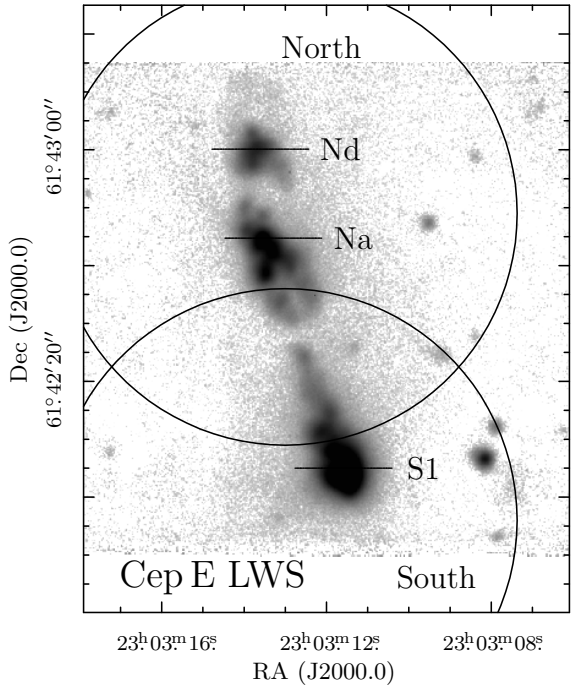}}} at 5.25 6.2
\put {\small RA (J2000.0)} at 5.25 0.5
\put {\footnotesize 23\fh 03\fm 16\fs} at 2.8 1
\put {\footnotesize 23\fh 03\fm 12\fs} at 5.35 1
\put {\footnotesize 23\fh 03\fm 08\fs} at 7.9 1
\put {\begin{sideways} \small Dec (J2000.0) \end{sideways}} at 0.5 6.2
\put {\begin{sideways} \footnotesize 61\fdg 42\arcmin 20\arcsec \end{sideways}} at 1 5.1
\put {\begin{sideways} \footnotesize 61\fdg 43\arcmin 00\arcsec \end{sideways}} at 1 8.75
\put {\large South} at 7.0 2.0
\put {\large North} at 5.2 10.3
\put {\Large Cep\,E LWS} at 3.5 2.0
\plot 4.8 3.7 6.3 3.7 /
\put {\large S1} at 6.8 3.7
\plot 3.7 7.3 5.2 7.3 /
\put {\large Na} at 5.7 7.3
\plot 3.5 8.7 5.0 8.7 /
\put {\large Nd} at 5.5 8.7
\endpicture
\caption{\label{cepe_h2} Positions of the apertures for LWS observations (circles)
in Cepheus\,E superimposed on an image in the 1\,--\,0\,S(1) line of molecular
hydrogen at 2.122\,$\mu$m. The slit positions of the three KSPEC spectra are also
indicated.}
\end{figure}

The NIR spectra were obtained in the period 26--29 August 1996 with 
the UH\,2.2-m telescope at the Mauna Kea Observatory. The KSPEC 
spectrograph \citep{hhiy94}, a cross-dispersed Echelle spectrograph, 
was employed. It is designed to provide medium-resolution spectra
in the 1\,--\,2.5\,$\mu$m region. A HAWAII 
1024\,x\,1024 detector array  was used. We observed at three different
positions in the outflow, as shown in Fig.\,\ref{cepe_h2}. Data reduction,
including flat-fielding, sky subtraction and extraction of the
spectra, was done using our own MIDAS routines. Absolute flux
calibration was not possible due to  non-photometric weather
conditions. The relative fluxes at wavelengths larger than 
$\lambda \, \mbox{\scriptsize $^>$} \!\!\!\!\mbox{\tiny $\sim$} \,
2.4$\,$\mu$m (1\,--\,0\,Q() lines) are unreliable because the
flat-field was quite poor. H$_2$ emission lines were detected in the H and
K band, which thus required wavelength calibration. For this purpose we 
used the OH night-sky emission lines and the tables of \citet{rlcmm00}.

\subsection{ISO data}

We employed three sets of ISO satellite data \citep{ksa_etal96}.
Cep\,E was observed at two different positions with the Long
Wavelength Spectrometer \citep{caa_etal96} in the LWS\,01 grating
mode. The apertures are indicated as circles in Fig.\,\ref{cepe_h2}
In the LWS\,01 mode, a spectrum from 43 to 197\,$\mu$m was 
obtained with a spectral resolution of about 200.
(See the ISO Handbook, Volume IV: LWS\,---\,The Long Wavelength
Spectrometer\footnote{
http://www.iso.vilspa.esa.es/manuals/HANDBOOK/IV/lws\_hb/} and
\citet{caa_etal96} for instrument and Astronomical Observing Template
details). 

In addition, the region was imaged  with the ISOCAM instrument
\citep{caa_etal96} through the CVF-filters in a wavelength range 
from 5 to 17\,$\mu$m and a pixel size of six arcseconds. We refer to 
the work of \citet{mnmtcs01}, where these data were first presented. 
The observed field  of the observations is larger than the NIR image 
shown in Fig.\,\ref{cepe_h2}. (See the ISO Handbook, Volume III: 
CAM -- The ISO 
Camera\footnote{http://www.iso.vilspa.esa.es/manuals/HANDBOOK/III/cam\_hb/} 
and \citet{caa_etal96} for instrument and AOT details.) A log of the ISO
observations used here is given in Table\,\ref{obslog}.

\begin{table}[t]
\caption{\label{obslog}Observation log for Cep.\,E ISO data.}
{\scriptsize
\begin{center}
\begin{tabular}{llccc}
TDT & Object & $\alpha$\,(J2000) &$ \delta$\,(J2000) & AOT  \\
\noalign{\smallskip}
\hline
\noalign{\smallskip}
56600912 & Cep\,E S & 23 03 13 & +61 41 56 & LWS01 \\
56601113 & Cep\,E N & 23 03 13 & +61 42 59 & LWS01 \\
79200740 & Cep\,E & 23 03 13 & +61 42 27 & CAM04 \\
\noalign{\smallskip}
\hline
\noalign{\smallskip}
\end{tabular}
\end{center}}
\end{table}

Data reduction was performed using ISO software (ISAP 1.6a and LIA 7.3 for the
LWS data and CIA for the ISOCAM-CVF data), and the data from standard pipeline
8.7 (LWS) and 10 (ISOCAM-CVF). In the LWS spectra glitches due to cosmic ray
hits were removed as well as the heavy fringing which occurred in the spectra.
Flux measurements were extracted using Gaussian fits to the lines plus  second
order polynomials for the background. Lines with a FWHM significantly smaller
than the instrumental profile are not considered as real, and are excluded from
further analysis.

\subsection{Submillimeter observations}

Submillimeter data  of the Cep\,E outflow were obtained in the 
$^{12}$CO(3\,--\,2) rotational line. The observations were carried out in the
period 18 -- 21 June, 1997 on the JCMT at Manua Kea.  We employed the B3 and C
heterodyne receiver. Data reduction was done using SPECX software. A spectrum
was measured at each point on a 7x11 grid with a grid point separation of six
arcseconds. We also obtained some data for the lines $^{12}$CO(4\,--\,3),
CS(7\,--\,6) and SiO(8\,--\,7) which we briefly describe below.

\section{Results}
\label{results}

\subsection{Near infrared data}
 
Almost 30 ro-vibrational transitions of molecular hydrogen were detected in the
H and K bands. These include the 1\,--\,0 S branch up to the 1\,--\,0\,S(9), 
the 1\,--\,0\, Q branch up to 1\,--\,0\,Q(8), the 2\,--\,1 S branch and even a 
few 3\,--\,2\,S branch lines. Some lines were not detected because they are
situated  in atmospheric absorption bands. Table\,\ref{cepekspeclines} provides
the complete  list of all relative fluxes detected with KSPEC. The line ratios
are quite similar in all three positions. As we will find, the K-band
extinction is quite high and might also have significant influence on the
measured excitation of the molecular hydrogen.

There are now 4 independent determinations of the flux ratio  R21 =
I(2--1~S(1))/I(1--0~S(1)), at various locations. For location Na, we find 
R21~=~0.12, in agreement with \citet{angcrbr00} (their North location). We
estimate  R21~$\sim$~0.1 from \citet{1997ApJ...474..749L}, while \citet{esdr96}
found R21~=~0.078$\pm$0.005 at location ND (roughly a 2\arcsec~ box).  Hence
the very low excitation reported in the latter work is not confirmed for this
or the other locations. For the location in the southern lobe, S1, we find
R21~=~0.13, while \citet{angcrbr00} find  R21~=~0.11. This is consistent with
typical 10\% errors in the 2--1 measurements.  

A subset of the  NIR emission lines detected here were also measured by
\citet{angcrbr00} for two positions, Na and S1, close to ours. This provides an
indirect  means of calibrating our data. The main problem, however, is to 
determine the total flux of the K-band lines within the ISO apertures in order
to directly compare and model the emission levels. The integrated 1--0~S(1)
lobe fluxes are 1.4~$\times$~10$^{-15}$~W~m$^{-2}$ (North) and
2.7~$\times$~10$^{-15}$~W~m$^{-2}$ (South) \citep{esdr96}. The flux ratio is
reasonably close to the value of 0.44  measured at the location of the peak
fluxes by \citet{angcrbr00}.

\begin{table}[t]
\caption{
Relative fluxes of H$_2$ lines in the H- and K-band measured with KSPEC in the
three observed slits (see Fig.\,\ref{cepe_h2}). The fluxes are normalised to
the 1\,--\,0\,S(1) line. The errors are about 10\,\% and 25\,\% (for $\lambda
\, \mbox{\scriptsize $^>$} \!\!\!\!\mbox{\tiny $\sim$} \, 2.4$\,$\mu$m).} 

\label{cepekspeclines} 
\centering
{\footnotesize
\begin{tabular}{lrrrr}
\noalign{\smallskip}
\hline
\noalign{\smallskip}
Transition & $\lambda_0$~~~ & Cep\,E & Cep\,E & Cep\,E \\[0mm]
 & [$\mu$m]~ & Na~~~ & Nd~~~ & S1~~~ \\[0mm]
\noalign{\smallskip}
\hline
\noalign{\smallskip}
1\,--\,0\,S(9) & 1.6873 &  0.025 &   0.014 &   0.035     \\
1\,--\,0\,S(8) & 1.7143 & $<$\,0.015 &   0.012 &   0.026     \\
1\,--\,0\,S(7) & 1.7475 &  0.097 &   0.093 &   0.144     \\
1\,--\,0\,S(6) & 1.7876 &  0.063 &   0.061 &   0.090     \\
1\,--\,0\,S(5) & 1.8353 &  0.261 &   0.293 &   0.322     \\
1\,--\,0\,S(4) & 1.8914 &  0.099 &   0.109 &   0.351     \\
1\,--\,0\,S(2) & 2.0332 &  0.328 &   0.323 &   0.349     \\
1\,--\,0\,S(1) & 2.1213 &  1.000 &   1.000 &   1.000     \\
1\,--\,0\,S(0) & 2.2226 &   0.339 &   0.309 &   0.281    \\
1\,--\,0\,Q(1) & 2.4059 &   1.101 &   1.010 &   1.106    \\
1\,--\,0\,Q(2) & 2.4128 &   0.417 &   0.379 &   0.341    \\
1\,--\,0\,Q(3) & 2.4231 &   1.301 &   1.160 &   1.215    \\
1\,--\,0\,Q(4) & 2.4368 &   0.367 &   0.323 &   0.340    \\
1\,--\,0\,Q(5) & 2.4541 &       0.584 &   0.648 &   0.779    \\
1\,--\,0\,Q(6) & 2.4749 &       0.276 &   0.164 &   0.209    \\
1\,--\,0\,Q(7) & 2.4993 &       0.548 &   0.430 &   0.464    \\
1\,--\,0\,Q(8) & 2.5270 &  $<$\,0.044 &        0.081 &   0.136    \\
2\,--\,1\,S(7) & 1.8523 &  $<$\,0.033 &   $<$\,0.010 &   0.051    \\
2\,--\,1\,S(4) & 2.0035 &  $<$\,0.016 &        0.015 &   0.040    \\
2\,--\,1\,S(3) & 2.0729 &       0.084 &        0.085 &   0.095    \\
2\,--\,1\,S(2) & 2.1536 &       0.043 &        0.041 &   0.048    \\
2\,--\,1\,S(1) & 2.2471 &       0.108 &        0.103 &   0.117    \\
2\,--\,1\,S(0) & 2.3550 &       0.019 &        0.016 &   0.027    \\
3\,--\,2\,S(6) & 2.0130 &  $<$\,0.011 &   $<$\,0.010 &   $<$\,0.007\\
3\,--\,2\,S(5) & 2.0650 &       0.016 &   $<$\,0.010 &        0.015  \\
3\,--\,2\,S(4) & 2.1274 &  $<$\,0.016 &   $<$\,0.010 &   $<$\,0.008 \\
3\,--\,2\,S(3) & 2.2008 &  $<$\,0.010 &        0.023 &   0.025   \\
3\,--\,2\,S(2) & 2.2864 &  $<$\,0.010 &        0.009 &   0.015  \\
3\,--\,2\,S(1) & 2.3858 &  $<$\,0.010 &        0.013 &   0.027 \\
\noalign{\smallskip}
\hline
\noalign{\smallskip}
\end{tabular}}
\end{table}


\subsection{ISO LWS data}

\begin{table}[t]
\renewcommand{\tabcolsep}{2pt}
\renewcommand{\baselinestretch}{1}
\caption{\label{flux} Observed lines in Cepheus\,E North and South. The H$_2$
fluxes measured in the ISOCAM data are co-added in the LWS apertures. The LWS 
fluxes are in $10^{-16}$\,W\,m$^{-2}$, the ISOCAM H$_2$ fluxes are in  
$10^{-15}$\,W\,m$^{-2}$.} 
\begin{center}
{\footnotesize
\begin{tabular}{llrrr}
Element & Transition & $\lambda_0$~~~~ & Cep\,E & Cep\,E \\[0mm]
  &   & $[\mu m]$~~ & North & South \\[0mm]
\noalign{\smallskip}
\hline
\noalign{\smallskip}
H$_2$ 	 &0--0\,S(7) 			&  5.510   & 4.2$\pm$0.8    & 3.4$\pm$0.7   \\[0mm]
H$_2$ 	 &0--0\,S(6) 			&  6.107   & 2.2$\pm$0.4    & 2.0$\pm$0.4   \\[0mm]
H$_2$ 	 &0--0\,S(5) 			&  6.908   & 8.3$\pm$1.7    & 5.3$\pm$1.1   \\[0mm]
H$_2$ 	 &0--0\,S(4) 			&  8.023   & 2.8$\pm$0.6    & 3.6$\pm$0.7   \\[0mm]
H$_2$ 	 &0--0\,S(3) 			&  9.662   & 6.1$\pm$1.2    & 6.6$\pm$1.3   \\[0mm]
H$_2$ 	 &0--0\,S(2) 			& 12.275   & 4.6$\pm$0.9    & 5.8$\pm$1.2   \\[0mm]
[OI]     &$^3$P$_1$\,--\,$^3$P$_2$        &63.184  &93.4$\pm$5.0    &129$\pm$6.5    \\[0mm]
o--H$_2$O&3$_{21}$\,--\,2$_{12}$          &75.380  &17.7$\pm$3.4    &$<$10.0~~~~~~~~\\[0mm]
o--H$_2$O&5$_{05}$\,--\,4$_{14}$          &99.492  &11.2$\pm$1.0    &$<$10.0~~~~~~~~\\[0mm]
CO       &25--24                          &104.445 &                &$<$10.0~~~~~~~~\\[0mm]
o--H$_2$O&2$_{21}$\,--\,1$_{10}$          &108.073 &                &               \\[0mm]
CO       &24--23                          &108.763 &
\raisebox{1.5ex}[-1.0ex]{$\big\}$$<$15.0} &
\raisebox{1.5ex}[-1.0ex]{$\big\}$$<$16.0} \\[0mm]
CO       &23--22                          &113.458 &                &               \\[0mm]
o--H$_2$O&4$_{14}$\,--\,3$_{03}$          &113.537 &
\raisebox{1.5ex}[-1.0ex]{$\big\}$$<$22.4} &
\raisebox{1.5ex}[-1.0ex]{$\big\}$$<$12.9} \\[0mm]
CO       &22--21                          &118.581 &$<$10.0~~~~~~~~~&$<$13.2~~~~~~~~\\[0mm]
CO       &21--20                          &124.193 &4.4$\pm$2.0     &$<$13.4~~~~~~~~\\[0mm]
p--H$_2$O&4$_{04}$\,--\,3$_{13}$          &125.353 &$<$8.0~~~~~~~~  &               \\[0mm]
p--H$_2$O&3$_{31}$\,--\,3$_{22}$          &126.713 &$<$6.0~~~~~~~~  &$<$6.5~~~~~~~~ \\[0mm]
CO       &20--19                          &130.369 &6.5$\pm$1.1     &10.3$\pm$1.8   \\[0mm]
CO       &19--18                          &137.196 &9.2$\pm$1.0     &13.3$\pm$2.5   \\[0mm]
p--H$_2$O&3$_{13}$\,--\,2$_{02}$          &138.527 &$<$2.5~~~~~~~~  &5.9$\pm$3.3    \\[0mm]
CO       &18--17                          &144.784 &12.3$\pm$1.5    &16.8$\pm$1.6   \\[0mm]
[OI]     &$^3$P$_0$\,--\,$^3$P$_1$        &145.525 &$<$2.0~~~~~~~~  &$<$3.0~~~~~~~~ \\[0mm]
CO       &17--16                          &153.267 &11.9$\pm$2.2    &19.8$\pm$1.0   \\[0mm]
[CII]    &$^2$P$_{3/2}$\,--\,$^2$P$_{1/2}$&157.741 &72.4$\pm$1.1    &86.1$\pm$2.0   \\[0mm]
CO       &16--15                          &162.812 &15.7$\pm$1.2    &20.7$\pm$2.2   \\[0mm]
CO       &15--14                          &173.631 &19.9$\pm$3.0    &30.9$\pm$5.8   \\[0mm]
o--H$_2$O&3$_{03}$\,--\,2$_{12}$          &174.626 &8.3$\pm$2.1     &14.0$\pm$0.9   \\[0mm]
o--H$_2$O&2$_{12}$\,--\,1$_{01}$          &179.527 &26.3$\pm$2.6    &29.1$\pm$2.2   \\[0mm]
CO       &14--13                          &185.999 &23.8$\pm$3.2    &28.9$\pm$3.5   \\[0mm]
\noalign{\smallskip}
\hline
\noalign{\smallskip}
\end{tabular}}
\end{center}
\end{table}

The ISO-LWS spectra of the two outflow lobes show a wide variety of atomic fine
structure and molecular lines. We detected rotational CO transitions from
J$_{up}$\,$=$\,14\,--\,21 and some water lines in both lobes. The CO emission
is on  average 1.5 times stronger in the southern outflow lobe, suggesting a
higher filling factor of the beam or a higher CO abundance. The same applies
for the H$_2$O lines. All detected lines and fluxes in the LWS spectra are
listed in Table\,\ref{flux}. The LWS spectra for both outflow lobes can be
found in Fig.\,2 of \citet{mnmtcs01}. We find no particular anomalies with
their derived fluxes.

\subsection{ISOCAM CVF data}

Pure rotational lines of the ground vibrational level of H$_2$ were detected 
(0\,--\,0\,S(2)\,..\,S(7)) with ISOCAM. Images in these lines as well as
spectra of selected pixels are presented in \citet{mnmtcs01}. To achieve our
aim, to simultaneously model the CO and H$_2$ lines, we co-add the spectra of
all pixels of the ICOCAM images which are situated in the LWS beam. Thus we
compare fluxes measured in the same aperture for both species, CO and H$_2$.
The co-added fluxes of the 0\,--\,0\,S lines in the two LWS apertures are given
in Table\,\ref{flux}.

\subsection{Submillimeter data}

\begin{figure}[t]
\resizebox{6.7cm}{!}{\includegraphics[angle=-90, bb=40 50 510 420]{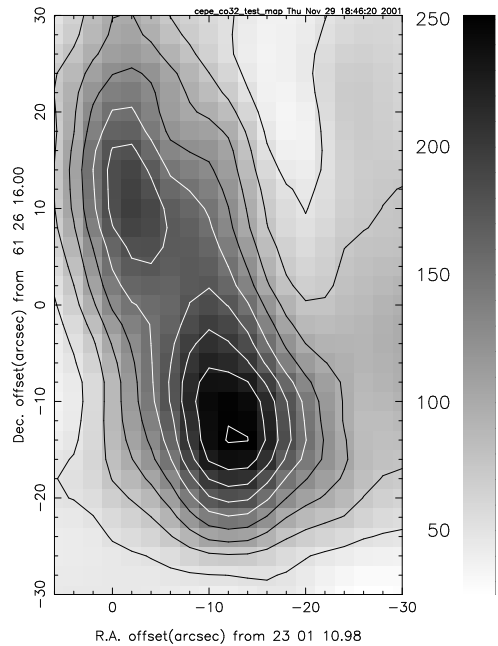}}
\caption{\label{cepe_gray} Gray-scale and contour map of the integrated flux in
the CO\,3\,--\,2 line of the Cepheus\,E outflow.}
\end{figure}

The two outflow lobes are also present in the CO(3\,--\,2) map 
(Fig.\,\ref{cepe_gray}). Note that the two lobes  are elongated in a direction
mis-aligned from the overall  outflow direction. At the brightest points, the
line shape shows signs of being optically thick (self absorption and
re-emission). High  velocity bullets with v$_{rel} \approx$\,100\,km\,s$^{-1}$
are also detected. In the CS(7\,--\,6) map we found weak emission only near the
source position and extremely weak emission at the three brightest knots of the
outflow (S1, Na and Nd). Nevertheless, the outflow structure is visible in this
line. At the three positions where the CO(4\,--\,3) line is observed we see the
same line structure as for the CO(3\,--\,2) line: optically thick in the line
centre and very fast CO bullets (v$_{rel} \approx$\,100\,km\,s$^{-1}$). The
northern positions observed in SiO(8\,--\,7) show no sign of a line. Only at
the two positions in the south very weak emission is detected.

\section{Extinction}
\label{extinct}

\subsection{Methods}

To determine the luminosity and excitation of a deeply-embedded outflow it is
necessary to remove the extinction. Extinction reduces observed fluxes from
their intrinsic values, especially for those from the shorter wavelength KSPEC
data. Hence, in order to test  shock models, we first adjust the KSPEC fluxes. 

Since the Cep\,E outflow is embedded in the parental cloud of the source (only
the southernmost part is visible at optical wavelengths as HH\,377), the
extinction is high. Moreover, the extinction may  vary not only over the field
of the LWS aperture but also through a lobe. That is, the K-band data will be 
more representative of  emission arising from low-excitation regions while ISO
and CO data  will sample more evenly. Here, however, since we do not have an
extinction map and cannot distinguish components of contrasting extinction, we
determine the extinction for each KSPEC location and then apply one value per
LWS and ISOCAM beam. 

Three means are at our disposal to estimate the extinction. First, 1.25\,mm
continuum emission from cool dust is  spatially coincident with the outflow
\citep{lel96}. This can be interpreted as an H column of
6~$\times$~10$^{22}$~cm in the  lobe locations, corresponding to 3--4~mag of
K-band extinction (incorrectly converted by \cite{lel96}). If half of this
emission lies on average in front of the lobe,  an extinction of 1.5--2~mag
would result. Submillimeter continuum observations yield a mean H$_2$ column of
3~$\times$~10$^{22}$cm within a 30$''$ radius, consistent with the above result
\citep{cwknrs01}.

Second, the Q-branch 1--0 lines beyond  2.4$\mu$m arise from the same upper
energy levels as the S-branch 1--0 lines within the K-band. Hence,
differential extinction between these wavelengths can, in theory, be determined
exactly since the number of emitted photons must be proportional to their
Einstein coefficients (radiative decay rates). Absolute extinction can then be
derived from the  differential value provided the properties of the dust are
known.

To apply this second method, we adopt a differential  extinction of the form 
$dlog(A_{\lambda}) = 0.4 A_K ((2.12\mu m)/\lambda)^{1.7} - 1.0)$. Using just
the 1\,--\,0\,S(1) and 1\,--\,0\,Q(3) lines (both originating  from the same
upper energy level), we determine K-band extinctions of A$_K \sim$ 2.6 --
3.2\,mag for the three  KSPEC slit positions. In contrast, the Q(2)/S(0) ratio
yields a uniform  extinction of just 1.1$\pm$0.1~mag. 

The Q-branch lines are, however, greatly effected by transmission and  are
often inaccurate. One good test for accuracy is to inspect the ratio of the
Q(3)/Q(1) lines which should be within  the range [0.91,1.12] if filter or
transmission distortions are low and if the gas is excited within the range
[1500K,3000K] \citep{1995A&A...296..789S}.  This was not case for the
previously derived values in  \citet{1997ApJ...474..749L} and \citet{mnmtcs01}.
\citet{mnmtcs01} find values of 0.58 and 0.86 for locations in the two lobes
and  \citet{1997ApJ...474..749L} find 0.78 and 0.88. Differential extinction
would only exacerbates the inconsistency. Here, we derive Q-branch Q(3)/Q(1)
ratios of 1.18, 1.16 and 1.10 for the 3 slits, which are comfortably consistent
when some differential extinction is taken into account. Nevertheless, even  in
this case, extinctions derived from the  S(1)/Q(3) and S(0)/Q(2) ratios do not
agree. 

\begin{figure*}[t]
\includegraphics[width=17.0cm]{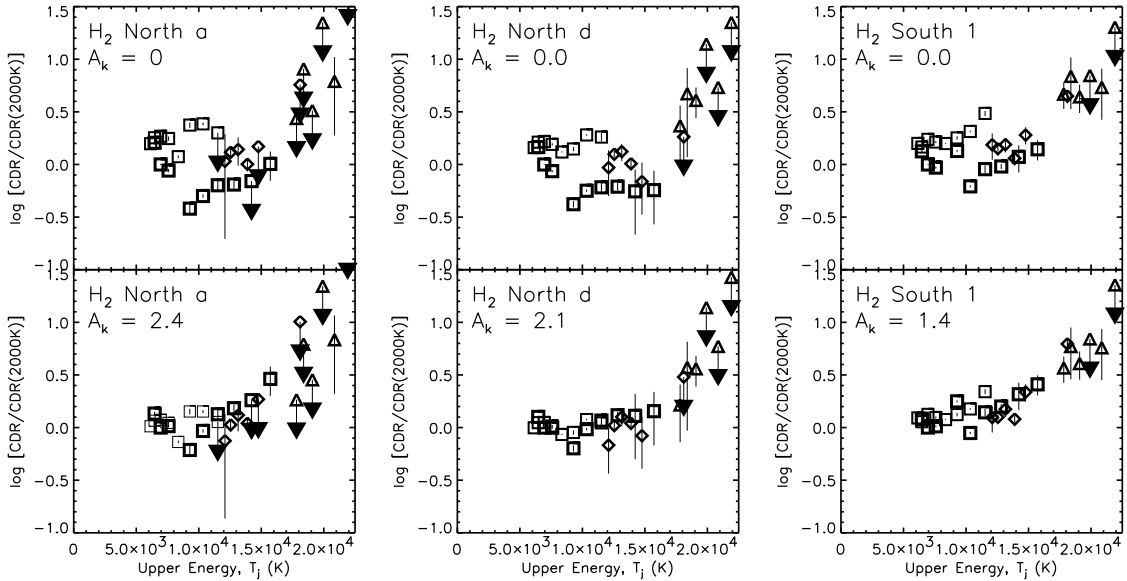}
\caption{\label{extinction} Column density ratios for the
near infrared H$_2$ ro-vibrational data, before applying extinction
(top panels) and after adjustment for extinction (lower panels). The 
slit locations and extinction are indicated on each panel. Upper bounds
are displayed as vertical lines. Squares represent 1--0 lines, with lighter
squares denoting the less-reliable columns derived from the Q-branch, 
diamonds indicated 2--1, and triangles indicate CDRs from 3--2 lines.} 
\end{figure*}

The third method also employs differential extinction but makes use of a large
fraction of the data. We plot the H$_2$ columns from all the levels across the
H and K bands and  then determine the extinction range which best correlates
the accumulated data. To achieve this result, however, one must apply the
normalised Column Density Ratio (CDR) method. Plotting the derived absolute
columns of H$_2$, N$_j$ against the upper energy level, E$_j$ hides the
information in the data since columns are spread out over three orders of
magnitude while error bars are only 10\% on many  data points.  In the CDR
method, the columns are normalised to the 1--0~S(1) flux and divided by the
equivalent LTE column of a gas at 2000K (see  \citet{fse02a} and
\citet{2000A&A...359.1147E} for details).  The results are shown in
Fig.~\ref{extinction}.

Fig.~\ref{extinction} confirms that high extinction is present.  Extinction is
responsible for the distribution of the data points along the edges of a
rhombus (top panels), which collapses to a line on applying the extinctions
indicated. The four sides of the rhombus consist of (i) the  K-band
1--0~S-branch, (ii) 1--0~Q-branch, (iii) 2--1~S-branch and (iv) H-band 
1--0~S-branch. We note no evidence for UV excitation and fluorescent emission
(as usually identifiable from rotational transitions within the higher
vibrational levels). Hence, the CDRs from the 2--1 lines should lie level or
below those from the 1--0 lines. However, we also require that the Q-branch
lines are not too far below the S-branch lines. In this manner, a good
approximation for extinction is reached in each case. 

\subsection{The intrinsic outflow}

The higher K-band extinction in the northern lobe by $\sim$ 0.7~mag implies
that this lobe is {\em not} intrinsically weaker than the southern
lobe in molecular hydrogen emission. The north-south 1--0 S(1) ratio was found
by \cite{esdr96} to be 0.54. This suggests  that the outflow may be much more
symmetric than appears.  Since the northern lobe is redshifted, this is also
consistent with  the lobe dynamics within a spherical cloud.

The mean density can be estimated on assuming the geometry of the  enveloping
cloud. Taking a uniform spherical cloud of radius 30$''$,  corresponding to
0.11\,pc, and an average column to the outflow of $\sim 3 \times
10^{22}$~cm$^{-2}$, yields a mean (H nuclei) density estimate of $1.3 \times
10^5 cm^{-3}$.

The extinction also implies that the intrinsic H$_2$ luminosity from the
1--0~S(1) line {\em alone} is 0.33~L$_\odot$, and the total H$_2$ luminosity 
will  probably lie in the range 3--9~L$_\odot$, depending on the type and
strength of  shock waves involved \citep{1995A&A...296..789S}. Notably, this
total is  very close to that estimated for the complete far-infrared cooling by
\cite{gnl01} of 2.8~L$_\odot$, which includes CO, H$_2$O, O and OH emission.

\begin{figure}[t]
\centering
\includegraphics[width=6.0cm, bb=160 4 423 408]{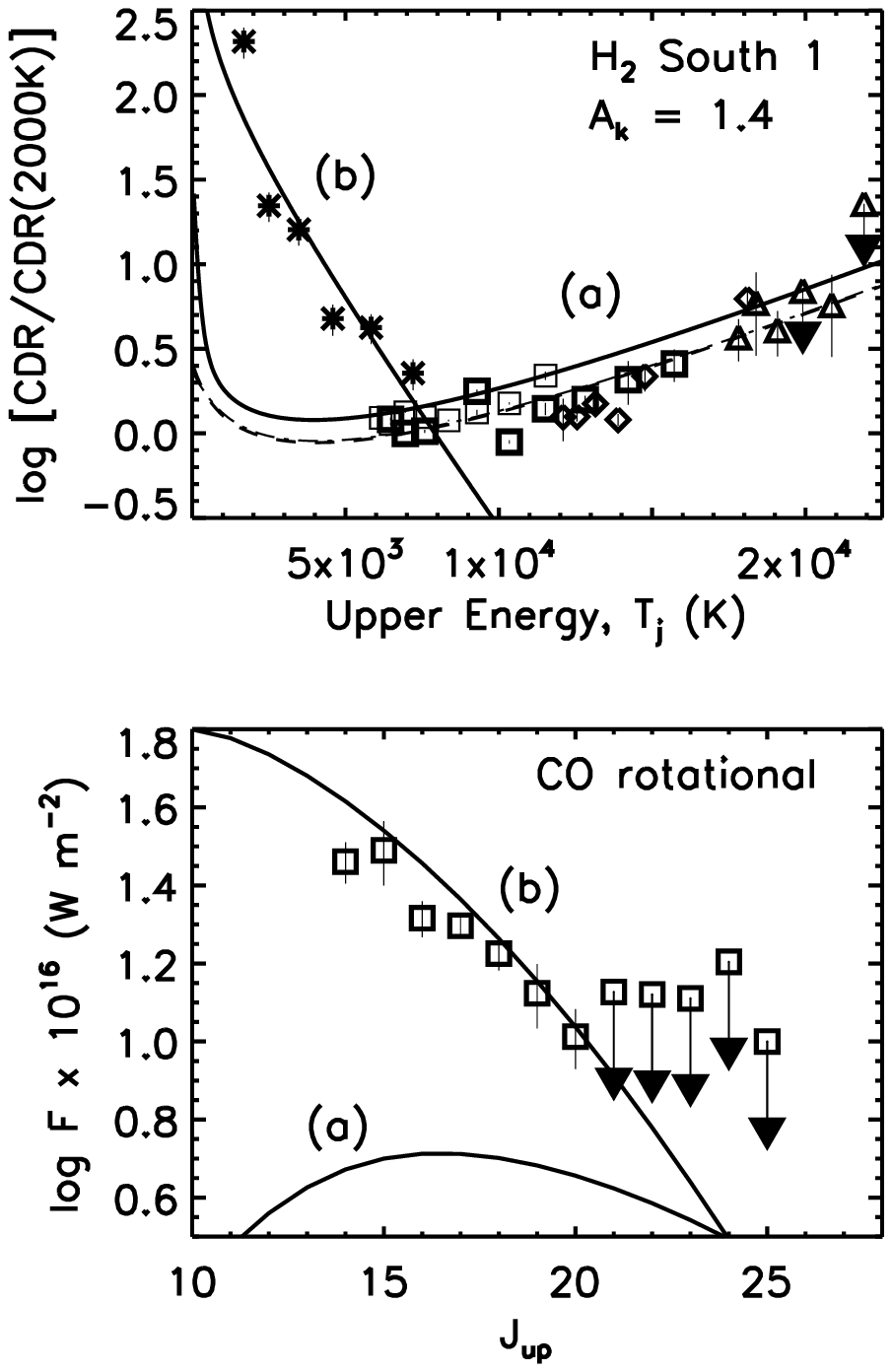}
\caption{\label{cepe-planar} Planar shock models for Cep\,E South. In addition
to the symbols described in Fig\,\ref{extinction}, the ISOCAM data are
represented by stars. {\bf (a)} A hot C-shock component with 32\,km\,s$^{-1}$
provides a fit to the  vibrationally-excited columns but does not contribute to
the CO rotational lines (other critical parameters are a density of
10$^5$~cm$^{-3}$, an Alfv\'en speed 1.5~km~s$^{-1}$, ion fraction
$\chi~10^{-6}$, a transverse magnetic field and
n(H$_2$)/(n(H)+2n(H$_2$))~=~0.3). The maximum temperature is 3160~K.  {\bf (b)}
A cool C-shock with  speed 11\,km\,s$^{-1}$ provides a  fit to the
rotationally-excited H$_2$ and CO but does not contribute to the vibrational
lines (other critical parameters are as above except 
n(H$_2$)/(n(H)+2n(H$_2$))~=~0.49998). The maximum temperature in the  shock is
950~K. Also modelled are abundances of $\epsilon(O)~=~4~10^{-4}$ and 
$\epsilon(C)~=~1.5~10^{-4}$.}
\end{figure}

\section{H$_2$ and CO Modelling}
\label{models}

To simultaneously model the KSPEC, ISOCAM and LWS CO data, we require a
calibration of the KSPEC data set. We employ the integrated 1--0~S(1) 
K-band fluxes of $1.45 \times 10^{-15}~Wm^{-2}$ (North) and 
$2.68 \times 10^{-15}~Wm^{-2}$ (South) tabulated by \cite{esdr96} 
since the associated areas correspond quite closely to the effective 
apertures of the ISO observations.

We have attempted to apply simple planar J-shock and C-shock models, without
success. We have employed an updated  shock code described in detail by
\cite{skd03}. The code assumes that a shock is stable and in a steady state,
the ion number is a conserved quantity and that the H$_2$ dissociation rate is
given by equilibrium conditions.  We take an ortho-para ratio for H$_2$ of 3.
Single J-type shock waves predict  high excitation H$_2$ spectra and C-type
shocks predict quite constant excitation across a wide range of upper energy
levels. 

Multiple shock waves are required, as demonstrated in Fig.~\ref{cepe-planar}.
Two carefully chosen C-type shocks provide a reasonable fit to the full set of
data. The top panel displays both the ISO H$_2$ data (stars at low T$_j$) and
the KSPEC data. The lower panel displays the CO rotational fluxes. It is not
clear, however, how two shocks with almost the same parameters can be found in
the three separate locations   since the excitation produced in planar shocks
is very sensitive to the shock velocity, field strength and ionization
fraction.

\begin{figure}
\centering
\includegraphics[width=6.0cm, bb=160 4 423 408]{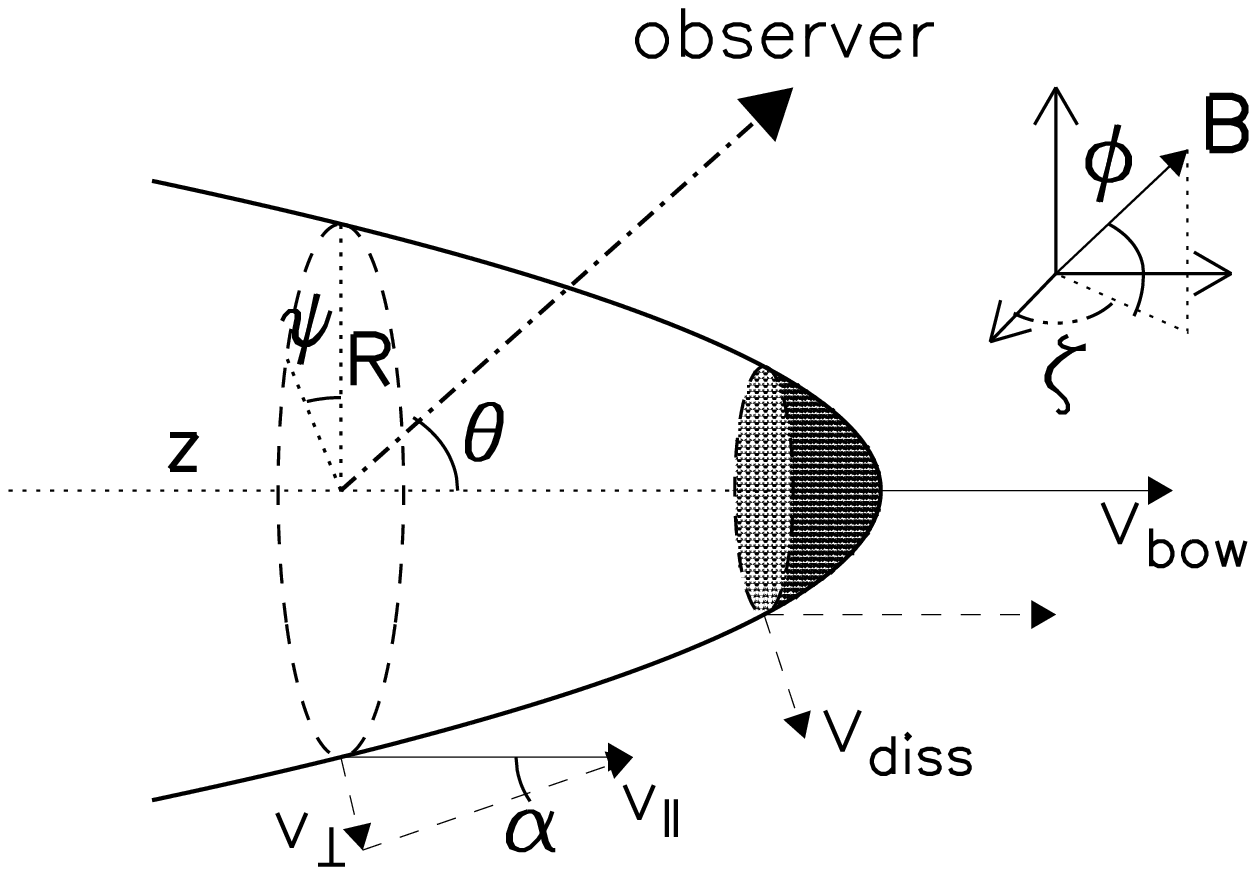}
\caption{\label{sketch} The geometrical parameters associated with a fast-moving
bow shock, as originally employed by \citet{sb90}. }
\end{figure}

Next, we fit a single paraboloidal C-type bow shock. C-type and J-type bow
shocks are modeled according to a scheme illustrated in Fig.\,\ref{sketch}.
Cylindrical  coordinates (z, R, $\phi$) are used, and the magnetic field B,
is defined via the density and Alfv\'en speed. The spectroscopic
results presented here are independent of the direction of the observer,
$\alpha$. The molecules are completely dissociated at the bow cap provided
the bow is moving faster than the appropriate dissociation speed limit.
The data provides a large number of constraints. For example, the CO fluxes
and excitation determine not only the CO abundance but also the density.
The excitation state of the vibrationally excited H$_2$ also determines
the density as well as the shape of the bow. The rotationally-excited H$_2$
determines the atomic hydrogen fraction.

\begin{figure}
\centering
\includegraphics[width=6.0cm, bb=160 4 423 408]{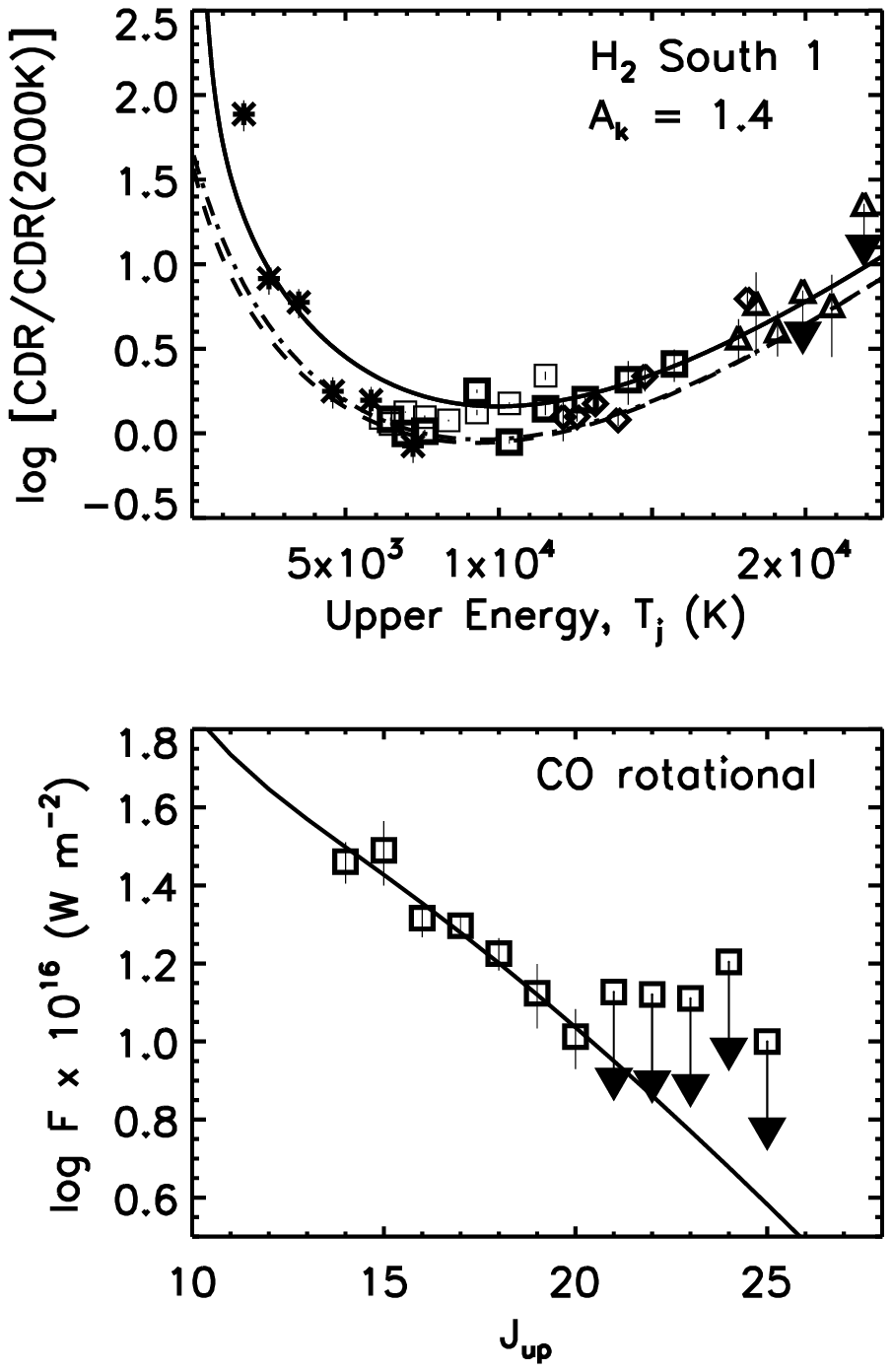}
\caption{\label{cbow-s1} A C-type bow shock model
for Cep\,E South 1. 
A C-type bow model with speed 120\,km\,s$^{-1}$,
a density of 10$^5$~cm$^{-3}$,
an Alfv\'en speed 1.5~km~s$^{-1}$, ion fraction $\chi~10^{-6}$ 
and n(H$_2$)/(n(H)+2n(H$_2$))~=~0.35, abundances of 
$\epsilon(O)~=~3~10^{-4}$ and $\epsilon(C)~=~1.4~10^{-4}$ 
(initially in CO) and a magnetic field aligned with the bow axis.
The H$_2$ model values are displayed for the ground (solid line ), first (dashed) and second
(dot-dash) vibrational levels.}
\end{figure}

\begin{figure}[t]
\centering
\includegraphics[width=6.0cm, bb=160 4 423 408]{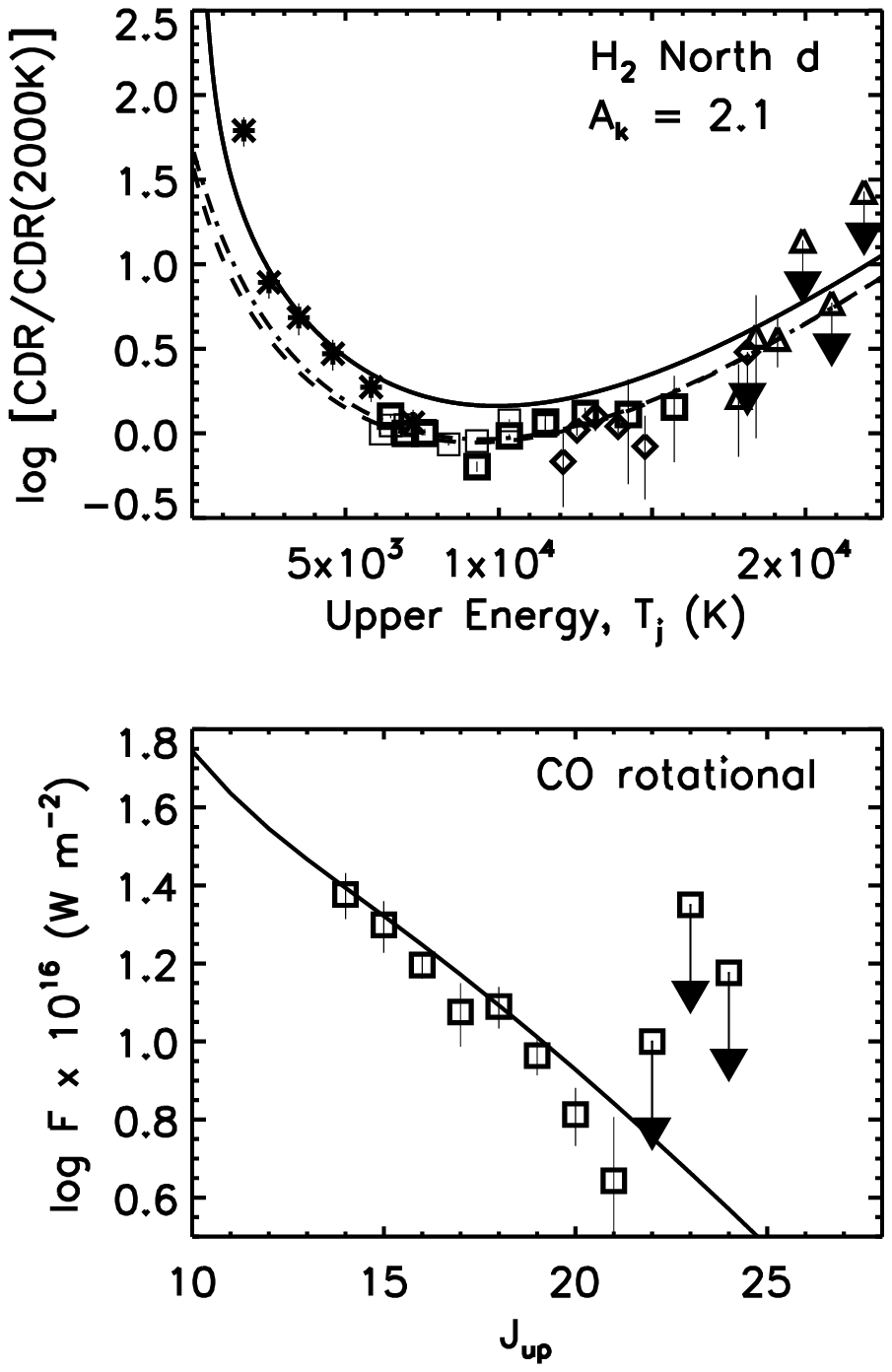}
\caption{\label{cbow-nd} A C-type bow shock model
for Cep\,E North (location Nd). The model bow
speed is again 120\,km\,s$^{-1}$ and critical parameters are as in 
Fig.~\ref{cbow-s1}, except for the initial abundance 
$\epsilon(CO) = 1.2 \times 10^{-4}$. Displayed lines
on the CDR diagram represent individual vibrational levels (see 
Fig.~\ref{cbow-s1}).}
\end{figure}

The final results are displayed in Fig.~\ref{cbow-s1} for Cep E South and
Fig.~\ref{cbow-nd} for Cep E North (location Nd for the KSPEC data). Note that
bow configurations are  strongly supported by two independent observations.
First, the  H$_2$ images display numerous bow-shaped structures in both
lobes  (e.g. Fig.~\ref{cepe_h2} and \cite{1997ApJ...474..749L} and, second, 
many locations within these 1--0 S(1) bows possess double-peaked  line 
profiles, as predicted by bow shock models \citep{1997IAUS..182...93E} and
numerical simulations \cite{1997A&A...318..595S}.


The displayed CBOW model fits are further evidence that these are bow shocks.
The ranges of critical parameters determined during the  modelling procedure
are given in Table~\ref{ranges}. Most significant  are (i) the density, (ii)
the molecular fraction, (iii) the bow shape  and (iv) the CO abundance. Less
critical parameters to the gas  excitation state are (i) the bow speed (the
excitation is fixed provided the bow speed exceeds a minimum value),  (ii) the
Alfv\'en speed (i.e. the magnetic field), (iii) the ion fraction (a minimum
value is necessary for a J-shock to be appropriate)  and (iv) the oxygen
abundance. The latter parameters influence the location of the molecular
emission along the bow surface but, provided the location exists and is not at
the apex, the excitation is not strongly affected. Perhaps the most significant
result is the atomic hydrogen  fraction of 0.1 -- 0.4. Additionally, the
derived CO abundances at all three locations lie close to $1 \times 10^{-4}$.

\begin{table*}[t]
\caption{Ranges for the critical parameters which describe the
observed shock excitation, as derived by comparing models to both the
H$_2$ and CO data. Location Nd is taken here. The criterion taken as a
good fit is a discrepancy of approximately 30\% between the majority of data points
and the model fluxes.  } 
\label{ranges} 
\centering
{\footnotesize
\begin{tabular}{llllllllll}
\noalign{\smallskip}
\hline
\noalign{\smallskip}
Model & Bow   & Density    & Magnetic & Alfv\'en & Ion     & $\epsilon$(C) & $\epsilon$(O) & bow  & H$_2$\\[0mm]
      & speed &  10$^5$ & field    & speed & fraction &   &   & shape& fraction\\[0mm]
      & km~s$^{-1}$ & cm$^{-3}$ & mG &  km~s$^{-1}$ &   10$^{-7}$&10$^{-4}$ &  10$^{-4}$ &      &        \\[0mm]
\noalign{\smallskip}
\hline
\noalign{\smallskip}

C-type & $>$~38 & 0.6--1.2 & 0.2--1.3 & 1.0--6.0 & 4--50  
                & 0.6--1.4 & 1.0--5.0 & 1.9--2.3 & 0.32--0.40  \\
J-type & $>$~17 & 2.0--5.0 & 0.3--0.6 & 1.5--3.7 &  --    
                & 0.3--0.6 & 1.3--7.0 & 1.4--1.6 & 0.14--0.26  \\

\noalign{\smallskip}
\hline
\noalign{\smallskip}
\end{tabular}}
\end{table*}

\begin{figure}
\centering
\includegraphics[width=6.0cm, bb=160 4 423 408]{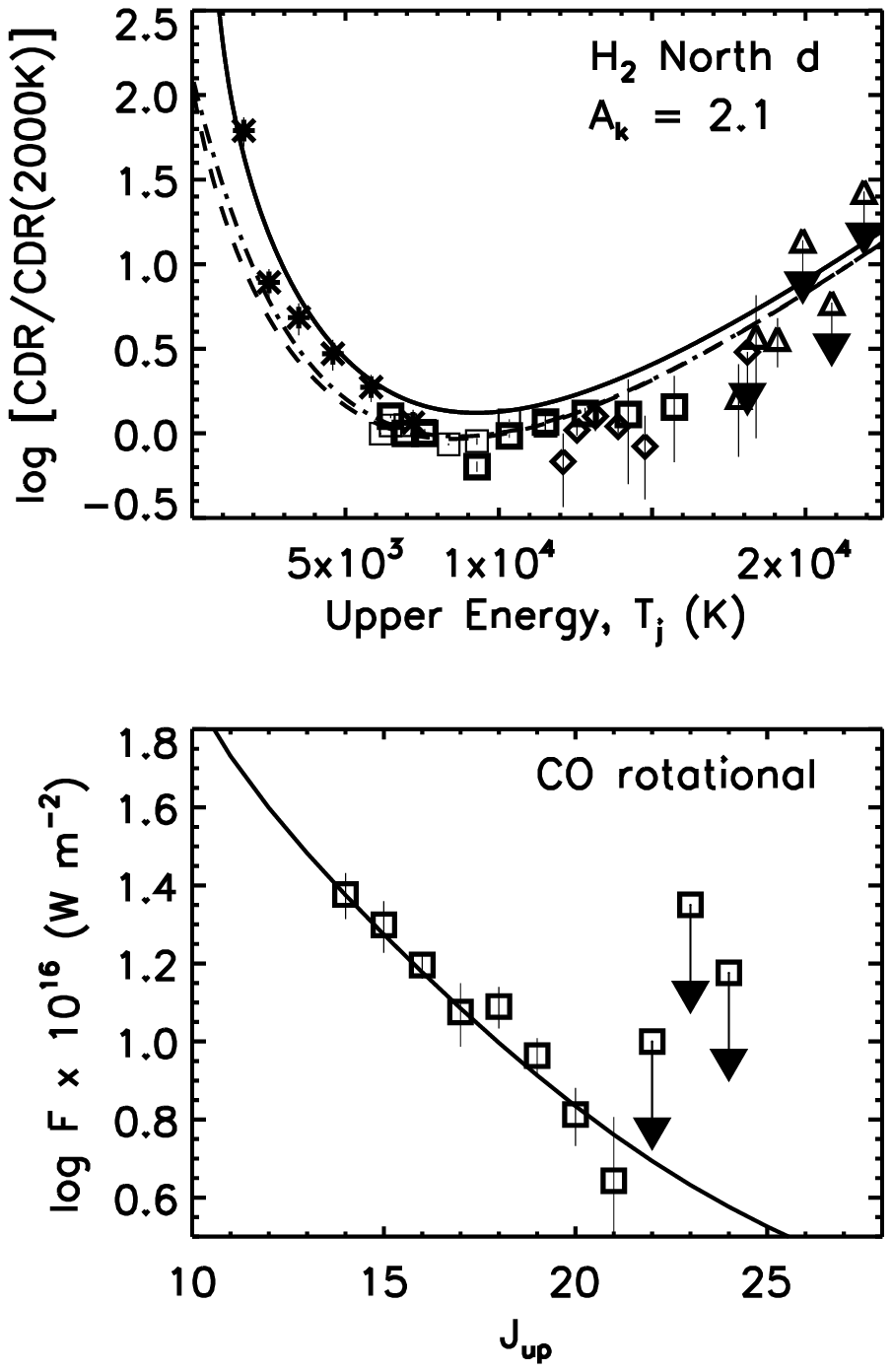}
\caption{\label{jbow-nd} A J-type bow shock model
for Cep\,E North, position Nd, with shape parameter s~=~1.5. 
A  speed of 60\,km\,s$^{-1}$, a density of 3~10$^4$~cm$^{-3}$,
an Alfv\'en speed 2.5~km~s$^{-1}$, n(H$_2$)/(n(H)+2n(H$_2$))~=~0.2, 
and abundances of $\epsilon(O)~=~4~10^{-4}$ and $\epsilon(C)~=~5~10^{-5}$ 
(initially in CO) and a magnetic field aligned with the bow axis were taken.
The H$_2$ model values are displayed for the ground (solid line ), first (dashed) and second
(dot-dash) vibrational levels.}
\end{figure}

Finally, we consider the type of J-type bow shock that could fit  the
spectroscopic data. Since J-shocks heat the gas impulsively, higher vibrational
levels are relatively well populated. Hence, to fit the data, a bow with longer
cooled wings than a paraboloid is needed to compensate. After varying the
critical parameters, Fig.~\ref{jbow-nd} shows that a bow with shape s~=~1.5
where  z/L = (1/s)\.(R/L)$^s$ fits the data extremely well. 

A major reason that quite blunt shapes are predicted here, in comparison to
those derived for Cepheus\,A, is that lower densities are involved. A lower
density is deduced from the relatively low CO fluxes in Cepheus E. The lower
density then  necessitates a higher fraction of hydrogen atoms to maintain a
small divergence from local thermodynamic equilibrium between the populations
of the H$_2$ vibrational levels.

\section{Further diagnostics}   

To distinguish between J and C-type bow models, we examine other signatures. 
We present the contributions to the cooling, calculated from the two best
fitting models to Cep E S1, in Table~\ref{bowfluxes}. The luminosities are
calculated by assuming a sufficient number of bow shocks to explain the H$_2$
1--0~S(1) luminosity.  Indeed, a reasonable number of distinct bow shocks are
inferred. We find from the above model that a single C-type bow of speed
v$_{bow}$ = 120~km~s$^{-1}$ and scale size L~=~10$^{15}$~cm emits
0.016~L$_\odot$ in the H$_2$ 1--0~S(1) line. Therefore, 20 such bows in each
lobe would suffice to provide the observed luminosity. Note that these  bows
would appear larger than L in size, with H$_2$ being thermally dissociated
until the normal component of velocity falls below 45~km~s$^{-1}$. For a
paraboloid, it is straightforward to show that this occurs at R~=~2.47~L and
z~=~3.05~L.

A single J-type bow with s~=~1.5, v$_{bow}$ = 60~km~s$^{-1}$ and scale size 
L~=~10$^{15}$~cm emits 0.02~L$_\odot$.  Hence, 16 such bows would be needed to
account for the H$_2$ emission. The apparent bow size is, however, larger. With
a dissociation speed limit of  25~km~s$^{-1}$, we find molecules survive only
for bow locations at R~$>$~4.76 ~L and z~$>$~4.36~L.

\begin{table}[t]
\caption{
Contributions to the intrinsic cooling of the flow in Cep E South
and the bow shock predictions. The total cooling estimates are extracted
from \cite{gnl01}. } 
\label{bowfluxes} 
\centering
{\footnotesize
\begin{tabular}{lrrr}
\noalign{\smallskip}
\hline
\noalign{\smallskip}
Coolant & Observed   &   C-Bow    &   J-Bow    \\[0mm]
        & L$_\odot$  &  L$_\odot$ & L$_\odot$  \\[0mm]

\noalign{\smallskip}
\hline
\noalign{\smallskip}

H$_2$ 1\,--\,0\,S(1) & 0.33      &  0.33      &   0.33   \\
H$_2$ dissociative   &  --       &  0.11      &   0.92   \\
H$_2$ rot+vibrat.    &  --       &  6.77      &   8.32    \\
CO rote              & $>$\,0.51 &  4.76      &  56.56    \\
CO rote (J\,$>$\,10) & 0.51      &  1.26      &   1.38    \\
CO vibe              &  --       &  0.01      &   0.03   \\
H$_2$O               & $>$\,0.67 &  1.53      &   3.71   \\
H$_2$O (FIR, T$>$800K) & 0.67    &  0.54      & 1.39     \\
OH                   & 0.17      &  0.01      &   0.01   \\
O{\small I} 63$\mu$m         & $>$\,0.21 &  1.72      &  21.36  \\
C{\small I} 370$\mu$m       &  --       &  0         &  0.17 \\

\noalign{\smallskip}
\hline
\noalign{\smallskip}
\end{tabular}}
\end{table}

The [O {\small I}] 63$\mu$m line is potentially a good shock diagnostic. 
However, the column density of hydrogen nuclei which provides
unit  optical depth at line centre is just $5 \times 10^{20}~cm^{-2}$ 
\citep{1989ApJ...342..306H}.
This may explain why the model line flux is overpredicted by factors of a
few (see Table\,\ref{bowfluxes}).
Given gas columns of $3 \times 10^{22}~cm^{-2}$, the
observed [O{\small I}] emission may well not have a shock origin. Furthermore, 
we  also detected  the [C{\small II}](158\,$\mu$m) line in the LWS spectrum.
The  measured ratio [O{\small I}](63\,$\mu$m)/[C{\small II}](158\,$\mu$m) is
about  1.3 (1.5) for the north and south lobes, respectively. In shocks, the
[C{\small II}](158\,$\mu$m) line is usually several orders of magnitute
fainter  than the [O{\small I}](63\,$\mu$m) \cite{1989ApJ...342..306H}. Hence
these lines may have a PDR origin, as already discussed by \citet{mnmtcs01}.
The  observed OH luminosity is, according to \cite{gnl01}, to be treated with
caution  since it may well be due to continuum pumping.

We conclude from Table\,\ref{bowfluxes} that the long wings of the J-type bow
produce extremely high fluxes in the lines from the cooler molecular gas. This
is inconsistent with all expectations since the total cooling then
exceeding the bolometric luminosity of the protostar.  Furthermore, in a
shock-dissipative momentum-driven outflow, we expect the mechanical power of
the flow to be very close to the total cooling. We estimate here the mechanical
power through CO J=2--1 observations by first noting that most of the emission
is from low-speed CO but most of the kinetic energy lies in the CO moving with
radial  speeds of 15--25~km~s$^{-1}$ \citep{1997A&A...323..223S} (a result of
the very flat line profile found for Cepheus E). Hence, given a blue lobe mass
of  0.16~M$_\odot$ \cite{1997ApJ...474..749L}, we derive a mechanical energy
of  $\sim 5 \times 10^{45}$~erg and a mechanical power $\sim 20~L_\odot$.
Although this estimate is subject to considerable uncertainty, it is consistent
with the C-type bow model.

A further diagnostic is provided by the integrated intensity of the
CO(3\,--\,2) emission shown in Fig.\,\ref{cepe_gray}. The C-bow model predicts 
a CO(3\,--\,2)/CO(18\,--\,17) ratio of 6.0 in contrast to the J-bow model
(276)  for the southern outflow lobe. The measured ratio is 4 (north) and 2
(south), much  nearer to the predictions of the C-type bow model.

The columns in the 0-0~S(3) and 0-0~S(5) appear lower than predicted in the
southern lobe. This suggest that the ortho-to-para ratio of H$_2$ may be under
three, the LTE value for high temperatures. This can occur in C-shocks since
the gas is gradually heated  from a low pre-shock temperature. Therefore, the 
gas maintains the  pre-shock ortho-para ratio. After some heating, however, 
atomic hydrogen can induce ortho-para conversion \citep{1997A&A...327.1206S}. 

Note that the distribution of shocks generated by a supersonic turbulent
velocity field  can produce an excitation that mimics that from a bow shock
surface \citep{2000A&A...359.1147E}. It can be shown that a power law number
distribution of shock speeds (with power law index $\alpha = -2s/(s-1)$ where
$N \propto v^{-\alpha}$) is required. This also  assumes that there is no
systematic magnetic field effects although, for  bow shocks with moderate
Alfv\'en speeds, we find that the magnetic field  direction has an
insignificant influence on the excitation. 

\section{Conclusions}

We have analysed the molecular outflow from the Cepheus\,E-MM source over a
broad wavelength  range, from the near infrared into the sub-millimeter regime.
We combined KSPEC NIR data, ISO  mid- and far-infrared spectra and JCMT sub-mm
observations. We have concentrated on the properties of the vibrationally and
rotationally excited  H$_2$ and CO since sufficient data points exist to permit
a {\em simultaneously} interpretation in terms of shock models using J- and
C-type physics. We investigated planar and curved shocks.

Cepheus E is quite exceptional in its strong radiative power (above
10~L$_\odot$) and short dynamical age (under 700~yr given an advance speed of
over 100~km~s$^{-1}$). Our main results are as follows:

\begin{enumerate}

\item Extinction is most accurately derived by satisfying multiple constraints
on the complete NIR data set rather than from specific line ratios. The
southern blueshifted lobe has a mean K-band extinction of 1.4~mag and the
northern redshifted lobe has  extinction in the range 2.1~--\,2.4~mag. This is
consistent with the extinction being internal to a spherical cloud, with a
density of 10$^5$~cm$^{-3}$ and radius $3 \times 10^{17}~cm^{-3}$.
Remarkably, the extinction-corrected H$_2$ columns, including the ISO data,
demonstrate a state close to local thermodynamic equilibrium.

\item We interpret all the measured H$_2$ and CO line fluxes simultaneously  in
terms of shock models. The best fitting models are obtained using {\it shock 
distributions} in the form of bow shocks.

\item C-type physics is strongly favoured mainly because J-type physics predicts
extremely strong emission from the low-excitation flanks of a bow, which is not
observed. Moreover, for the J-shock model, the  outflow power and momentum become 
implausibly large in comparison to the values found for
the protostar  and the bipolar outflow, respectively. 

\item High resolution H$_2$ 1\,--\,0 S(1) imaging has shown that the lobes consist
of numerous emission knots of size $1--3 \times 10^{15}$~cm \citep{1997ApJ...474..749L}.
We find that of order 20  such bow shocks, close to paraboloidal in shape, are 
required by the model to explain the intergrated emission for each lobe.

\item  No significant differences in the density or abundances
are found between positions. We also find no significant differences 
in the H$_2$ excitation when analysing the integrated  spectra of a 
whole outflow lobe or a
6\arcsec\,$\times$\,6\arcsec\, pixel-sized part. This is  in agreement with
\citet{esdr96} and suggests that the large bows are built up of  smaller
unresolved bow shocks, generated by flow instabilities.

\item The pre-shock medium is not fully molecular in these models. We have
found a mean atomic fraction of $n(H)/(n(H)+2n(H_2)) =0.4$. With bow speeds of
120~km~s$^{-1}$, extensive bow apices are predicted within which molecules are
completely destroyed. The reformation time is of order 10$^{17}$/n$_c$(H),
where n$_c$(H) is the atomic density in the compressed layers. Hence, a
reformation time of $\sim$ 3000~yr may be achieved for
n$_c$(H)~=~10$^6$~cm$^{-3}$.

\item The total mass set into motion by the outflow is low, 0.25\,M$_{\odot}$
\citep{1997ApJ...474..749L}. We estimate a total cloud mass of 5\,M$_{\odot}$
given a density of 10$^5$~cm$^{-3}$ and radius 2.2\,$\times$\,10$^{17}$~cm.
With an outflow half-opening angle of $\beta$, a fraction f$_c$ =
1\,-\,cos\,$\beta$ of the cloud would be disturbed by two lobes. Taking
$\beta$~=~20$^\circ$, yields  f$_c$~=~0.06, and thus provides a consistent
basic model. \end{enumerate}

High outflow powers have been uncovered from many other Class 0
protostars. Mechanical luminosity exceeds 50\% of the bolometric luminosity for
all of the Perseus Class 0 sources \citep{1998ApJ...509..733B}. On
the other hand, a few Class 0 protostars such as L1527 and B335 possess 
relatively weak outflows as measured by total far-infrared luminosities
\citep{gnl01} and CO momentum flow rates \citep{1996A&A...311..858B}. In this
respect, the Cepheus E outflow could represent a powerful but abrupt
evolutionary phase, about to be brought to a halt as underlying jets exit a
compact 0.1~pc cloud. 

\begin{acknowledgements}
Jochen Eisl\"offel and Dirk Froebrich received financial support from the DLR
through Verbundforschung grant 50\,OR\,9904\,9. The ISO Spectral Analysis
Package (ISAP) is a joint development by the LWS and SWS Instrument Teams and
Data Centers. Contributing institutes are CESR, IAS, IPAC, MPE, RAL and SRON.
\end{acknowledgements}

\end{document}